\documentclass[twocolumn,nofootinbib,floatfix,
superscriptaddress]{revtex4}        
\usepackage{graphicx}
\usepackage{epsfig}
\usepackage{bm}
\usepackage[T1]{fontenc}
\usepackage[latin9]{inputenc}
\usepackage{amssymb}
\usepackage{float}
\usepackage{amsmath}
\usepackage{dcolumn}
\usepackage{cancel}
\usepackage[colorlinks]{hyperref}
\usepackage[usenames,dvipsnames]{color}
\hypersetup{
     breaklinks=true,
    pdfstartview={FitH},    
    colorlinks=true,       
    linkcolor=blue,          
    citecolor=red,        
    filecolor=magenta,      
    urlcolor=blue,           
    anchorcolor=green,      
    linktocpage=true
}

\newcommand{\be}{\begin{equation}}
\newcommand{\ee}{\end{equation}}
\newcommand{\De}{\Delta}

\begin{document}

\title{Barrow holographic dark energy in the Brans-Dicke cosmology}

\author{S. Ghaffari}
\email{sh.ghaffari@riaam.ac.ir}
\affiliation{Research Institute for Astronomy and Astrophysics of Maragha (RIAAM), University of Maragheh, P.O. Box 55136-553, Maragheh, Iran}

\author{G.~G.~Luciano}
\email{giuseppegaetano.luciano@udl.cat}
\affiliation{Applied Physics Section of Environmental Science Department, Universitat de Lleida, Av. Jaume
II, 69, 25001 Lleida, Catalonia, Spain}

\author{S.~Capozziello}
\email{capozziello@na.infn.it}
\affiliation{Dipartimento di Fisica ``E. Pancini'', Universit\`a di Napoli ``Federico II'', Via Cinthia 9, I-80126 Napoli, Italy}
\affiliation{Scuola Superiore Meridionale, Largo S. Marcellino 10, I-80138 Napoli, Italy}
\affiliation{Istituto Nazionale di Fisica Nucleare (INFN),
Sez. di Napoli, Via Cinthia 9, I-80126 Napoli, Italy}

\date{\today}

\begin{abstract}
We construct a holographic model for dark energy 
in the Brans-Dicke cosmology by using the 
holographic principle considering the Barrow entropy instead of the standard Bekenstein-Hawking one. 
The former arises from the effort to 
account for quantum-gravitational effects in black-hole physics and, according to the  gravity-thermodynamic conjecture, in the cosmological framework. In order to explore the cosmological
consequences of our model, we consider the Hubble horizon
as the IR cutoff. We investigate both the non-interacting and
interacting cases with the sign-changeable and linear interactions, showing that they can explain the present accelerated phase of the Universe expansion, in contrast to the standard Holographic Dark Energy model.
We then perform the  stability analysis according to 
the squared sound speed. We find that, whilst the non-interacting 
model is unstable against  small perturbations, 
the sign-changeable interacting one can be stable only for suitable values of the model parameters. On the other hand, the linear interacting 
model always predicts a stable Universe. The consistency of the model
with respect to cosmological observations is discussed. 
\end{abstract}
 \maketitle

\section{Introduction}
\label{Intro}
The recent technological progress  has ushered  a new era
for observational cosmology, the so-called  {\it Precision Cosmology} \cite{Primack:2006it}
allowing for significant advances
in our understanding of the Universe. Yet, most of the Universe matter-energy
composition remains elusive. 
It is a fact that about 25 percent of matter content consists of an enigmatic substance - the Dark Matter (DM)~\cite{Zwicky:1933gu} - the existence of which is established by astrophysical observations at large scales. Possibly even more mysterious is the Dark Energy (DE) component (roughly 70 percent), which is invoked to account for the current accelerating expansion of the Universe. 
In spite of strong evidences from Supernova SNIa observations,
baryon acoustic oscillations and gravitational waves,  
a complete explanation for the origin of DE is an open issue~\cite{Nojiri:2006zh,Bamba:2012cp}. Undoubtedly, this 
reveals the necessity of novel physics to be solved.

So far, several theoretical approaches to DE problem 
have been developed in the literature. The possibility 
that DE could be modeled 
by the cosmological constant acting as source
of the vacuum energy was first
considered as a natural way out of the DE puzzle. 
Nevertheless, this scenario seems to clash with 
our field theoretical understanding of the quantum properties of  vacuum.
Alternative explanations were later offered by 
modified gravity theories based on the introduction of either
geometric terms or extra degrees of freedom
in the Einstein-Hilbert action (see~\cite{Copeland:2006wr,Capozziello:2003tk,Capozziello:2002rd,Nojiri:2010wj,Capozziello:2011et,Clifton:2011jh,Joyce:2014kja} for  reviews).  
Among these theories, the Brans-Dicke (BD) gravity 
is an extension of General Relativity built upon the replacement
of the gravitational coupling constant $G$ by the inverse
of a varying scalar field $\phi$~\cite{Weinberg}. The first motivation for this choice was to achieve a theory more in agreement with the Mach Principle than GR \cite{Brans, Faraoni}. While being able to describe 
the accelerating expansion of the current Universe, 
this model predicts a value for the BD ($\omega$) parameter that is  
lower than the observational limit~\cite{Banerjee:2000mj,Acquaviva:2007mm},  thus requiring to use different DE sources~\cite{Capoz96,Gong:1999ge,Banerjee:2000gt,Gong:2004,Lamb,Kim:2005gk,Setare:2006yj,Xu:2008sn,Khodam-Mohammadi:2014wla,Kazem} to be reconciled with phenomenology.

In this respect, a useful paradigm is given by the Holographic Dark Energy (HDE)~\cite{Cohen:1998zx,Horava:2000tb,Thomas:2002pq,Li:2004rb,pnp,Bamba:2012cp,Ghaffari:2014pxa,Wang:2016och}. In the primary formulation, such an approach provides a precise quantitative analysis of DE, originating from the holographic principle with Bekenstein-Hawking (BH) entropy and the Hubble horizon as its IR cutoff. However, the shortcomings in describing the history of a flat Friedmann-Robertson-Walker (FRW) Universe have motivated some tentative changes of this model. For instance, HDE has been employed to solve the DE problem in the BD framework~\cite{Gong:1999ge,Lamb,Kim:2005gk,Setare:2006yj,Banerjee:2000gt,Gong:2004,Xu:2008sn,Khodam-Mohammadi:2014wla} by considering different IR cutoffs~\cite{Wang:2016och,Ghaffari:2014pxa,Xu:2008sn} or suitable interactions with DM~\cite{Jamil:2010ed}. 

On the other hand, modified HDE models based on deformations of the entropy area law have been proposed in~\cite{Tavayef:2018xwx,Saridakis:2018unr,Saridakis:2020zol,Nojiri:2019skr,Moradpour:2020dfm,Drepanou:2021jiv,Nojiri:2021jxf,Fara,Nojiri:2021iko,yangluc}, while still keeping the Hubble radius as IR cutoff. Due to the complex (and mostly unknown) thermostatistical properties of gravity systems, various deformed entropies have been adopted~\cite{Tsallis:1987eu,Kaniadakis:2002,Barrow:2020tzx}, all of which recovering BH entropy in a certain limit.  In this context, a promising framework is provided by the Barrow Holographic Dark Energy (BHDE)~\cite{Saridakis:2020zol}, 
which relies on a quantum-gravitational deformation of BH area law in the form~\cite{Barrow:2020tzx}
\be
\label{BE}
S_\De\,=\,\left(\frac{A}{A_0}\right)^{1+\Delta/2}\,,
\ee
where $A$ is the standard horizon area and $A_0$ the Planck area. 
Quantum gravity effects are parameterized by the Barrow exponent $0\le\Delta\le1$, with $\Delta=0$ giving the BH
limit, while $\Delta=1$ corresponding to the maximal entropy
deformation. It is worth  mentioning that cosmological constraints on $\Delta$ have been inferred in~\cite{Anagnostopoulos:2020ctz,Leon:2021wyx,Barrow:2020kug,Jusufi:2021fek,Dabrowski:2020atl,Saridakis:2020cqq,Luciano:2022pzg,Vagnozzi:2022moj,LucThermal1,LucThermal2}. The possibility for a running $\Delta$ has also been discussed in~\cite{DiGennaro:2022ykp}.

BHDE is inspired by similar generalizations of HDE based on Tsallis~\cite{Tavayef:2018xwx,Saridakis:2018unr,Nojiri:2019skr} and Kaniadakis~\cite{Drepanou:2021jiv} entropies.  At a first glance, it appears as a
proper model to trace the evolution of the Universe in the standard cosmological framework. Just as Tsallis Holographic Dark Energy, however, BHDE suffers
from instability problems. Thus, according to the dynamic behavior
of HDE and its modified-entropy based extensions, 
it could be more appropriate to study the cosmological features of BHDE in extended frameworks and, in particular, in the
BD dynamic framework~\cite{Setare:2006yj,Jamil:2010ed}. In \cite{Xu:2008sn}, the authors have shown that the non-interacting HDE with the Hubble radius as IR cutoff cannot explain the current accelerated expansion of Universe in the BD theory.
While in a similar analysis, it has been shown that the Tsallis and Kaniadakis Holographic Dark Energy, due to the generalized  entropy, with the Hubble cutoff in the BD gravity,  can lead the accelerated Universe phase, even in the absence of interaction between two dark sectors~\cite{Ghaffari:2018,KBD}. 

Starting from the above premises, in what follows we 
examine the implications of using BHDE to model dark energy in BD
cosmology. To this goal, we resort to the holographic principle 
with Eq.~\eqref{BE} as horizon entropy and the Hubble radius as cutoff. 
In the next section, we extract the BHDE
density in BD gravity and investigate the evolution of its cosmological parameters
for a spatially flat Universe in the non-interacting case. The same study is developed in Sec.~\ref{Inter} for the sign-changeable and linear interacting models. Stability analysis is investigated in Sec.~\ref{Stability}. Conclusions and outlook are finally summarized in Sec.~\ref{Conc}.

\section{Non-interacting Barrow holographic dark energy in the Brans-Dicke cosmology}
Throughout this work we consider a homogeneous and isotropic Friedman-Robertson-Walker
(FRW) geometry with metric
\begin{equation}
{\rm d}s^2=-{\rm d}t^2+a^2(t)\left(\frac{{\rm d}r^2}{1-kr^2}+r^2{\rm
d}\Omega^2\right),\label{metric}
\end{equation}
where $k=0,1,-1$ represent a flat, closed and open maximally
symmetric space, respectively. 

The Brans-Dicke field equations for a spatially flat FRW Universe which is filled by
 pressureless DM and DE are~\cite{Banerjee:2000gt,Gong:2004,Kim:2005gk,Setare:2006yj,Xu:2008sn}
\begin{equation}\label{Friedeq01}
\frac{3}{4\omega}\phi^2H^2-\frac{\dot{\phi}^2}{2}+\frac{3}{2\omega}H\dot{\phi}\,\phi=\rho_M+\rho_\Lambda,
\end{equation}
\begin{equation}\label{Friedeq02}
\frac{-\phi^2}{4\omega}\left(\frac{2\ddot{a}}{a}+H^2\right)-\frac{1}{\omega}H\dot{\phi}\,\phi
-\frac{1}{2\omega}\ddot{\phi}\,\phi-\frac{\dot{\phi}^2}{2}\left(1+\frac{1}{\omega}\right)=p_\Lambda,
\end{equation}
The scalar field evolution equation is
\begin{equation}\label{motiom eq}
\ddot{\phi}+3H\dot{\phi}-\frac{3}{2\omega}\left(\frac{\ddot{a}}{a}+H^2\right)\phi=0.
\end{equation}
where $\omega$ is the BD parameter and $\phi^2=\omega/(2\pi G_{eff})$ is the BD scalar field,
with $G_{eff}$ being the effective gravitational constant.
Also, $H=\dot{a}/a$ is the Hubble parameter and $\rho_m$, $\rho_\Lambda$ and $p_\Lambda$ 
are the pressureless DM density, DE density and pressure of DE, respectively. 
Following~\cite{Banerjee:2000gt,Gong:2004,Kim:2005gk,Setare:2006yj,Xu:2008sn}, 
we consider the BD field as a power law of the scale factor, $\phi \propto a^n$. Then we have
\begin{eqnarray}\label{phidot}
\dot{\phi}&=&nH\phi,\\[2mm]
\ddot{\phi}&=&n^2H^2\phi+n\dot{H}\phi.
\end{eqnarray}
where dot denotes derivative with respect to the cosmic time.
The energy conservation equations for non-interacting DE and matter are
\begin{equation}
\dot{\rho}_D+3H\left(1+\omega_D\right)\rho_D=0,\label{ConserveDE}
\end{equation}
and 
\begin{equation}
\dot{\rho}_m+3H\rho_m=0,\label{ConserveCDM}
\end{equation}
where $\omega_D=\frac{p_D}{\rho_D}$ denotes the equation of state (EoS) parameter of DE.
With the definition of the critical energy density, $\rho_{cr}=\frac{3\phi^2H^2}{4\omega}$,
the dimensionless density parameters can be written as
\begin{eqnarray}\label{Omega}
\Omega_m=\frac{\rho_m}{\rho_{cr}}&=&\frac{4\omega\rho_m}{3\phi^2H^2},\\[2mm]
\Omega_D=\frac{\rho_D}{\rho_{cr}}&=& \frac{4\omega\rho_D}{3\phi^2H^2}. 
\label{Omega2}
\end{eqnarray}
Using the above definitions~(\ref{Omega}) and ~(\ref{Omega2}), one can rewrite the Friedman Eq.~(\ref{Friedeq01}) as follows
\begin{eqnarray}\label{OmegaT}
\Omega_m+\Omega_D=1+ 2n-\frac{2\omega n^2}{3}.
\end{eqnarray}
Here, by using the  holographic principle with the generalized Barrow entropy~\eqref{BE}, we can construct the Barrow holographic dark energy (BHDE) in BD gravity, 
in which the apparent horizon in the flat Universe is considered as the IR cut-off $(L=H^{-1})$,
as follows
\begin{equation}\label{rho1}
\rho_D=B\phi^{2+\Delta}H^{2-\Delta}.
\end{equation}
Here, $B$ is an unknown constant. 
In the limit of Einstein gravity, where $G_{eff} \rightarrow G$,
the BD scalar field becomes trivial ($\phi^2=\frac{\omega}{2\pi G}=4\omega M_p^2$) and 
the BHDE in standard cosmology can be recovered \cite{Srivastava}.
Also for $\Delta\rightarrow 0$ and $G_{eff} \rightarrow G$,
Eq.~(\ref{rho1}) reduces to the energy density of the original HDE in standard cosmology~\cite{Xu:2008sn}.

Taking the time derivative of Eq.~(\ref{rho1}), we have
\begin{equation}\label{rhodot}
\dot{\rho}_D=\left(2+\Delta\right)nH\rho_D+\left(2-\Delta\right)\frac{\dot{H}}{H}\rho_D.
\end{equation}
By differentiating Eq.~(\ref{Friedeq01}) and combining the result with 
Eqs.~(\ref{ConserveCDM}),~(\ref{Omega}),~(\ref{Omega2}),
~(\ref{phidot}) and~(\ref{rhodot}), we obtain
\begin{eqnarray}\label{Hdot1}
\frac{\dot{H}}{H^2}&=&\Big[-\frac{9\Omega_m^0H_0^2\phi_0^2}{4\omega \phi^2 H^2}+(2\omega n^2-6n-3)\frac{n}{2\omega}\nonumber\\[2mm]&+&\,
nB(2+\Delta)\left(\frac{\phi}{H}\right)^\Delta\Big]\nonumber\\[2mm]&\times&
\Bigg[B(\Delta-2)\left(\frac{\phi}{H}\right)^\Delta+\frac{3}{2\omega}-n^2+\frac{3n}{\omega}\Bigg]^{-1},
\end{eqnarray}
where the zero-subscript represents present time.
Using Eqs.~(\ref{ConserveDE}) and~(\ref{rhodot}), 
we find the EoS parameter for the BHDE model in the BD gravity as
\begin{eqnarray}\label{EoS1}
\omega_D=-1-\frac{\left(2+\Delta\right)n}{3}+\frac{\Delta-2}{3}\frac{\dot{H}}{H^2}.
\end{eqnarray} 
For $\Delta= 0$ and  $n=0$,
this equation coincides with the original HDE in the BD gravity~\cite{Xu:2008sn}.
Also in the limiting case  $n=0$ $(\omega\rightarrow\infty)$,  
the EoS parameter for BHDE in standard cosmology is restored  \cite{Srivastava}.

In the following, using the accelerating parameter relation $q=-1-\frac{\dot{H}}{H^2}$ and Eq.~(\ref{Hdot1}),
the deceleration parameter can be extracted as
\begin{eqnarray}\label{q1}
q&=&-1+\Big[\frac{9\Omega_m^0H_0^2\phi_0^2}{4\omega \phi^2 H^2}-\left(2\omega n^2-6n-3\right)\frac{n}{2\omega}\nonumber\\[2mm]&-&\,
nB\left(2+\Delta\right)\left(\frac{\phi}{H}\right)^\Delta\Big]\nonumber\\[2mm]&\times&
\left[B\left(\Delta-2\right)\left(\frac{\phi}{H}\right)^\Delta+\frac{3}{2\omega}-n^2+\frac{3n}{\omega}\right]^{-1}
\end{eqnarray}
\subsection{Cosmological Evolution} 
\begin{figure}[htp]
\begin{center}
\includegraphics[width=8.7cm]{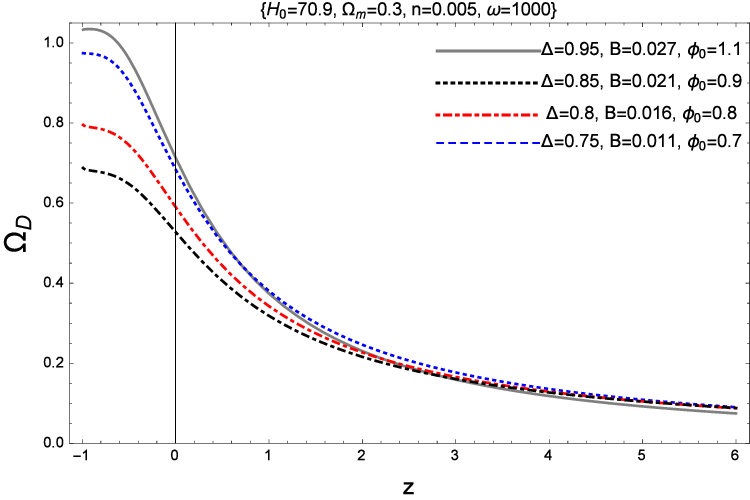}
\vspace{1mm}
\includegraphics[width=8.7cm]{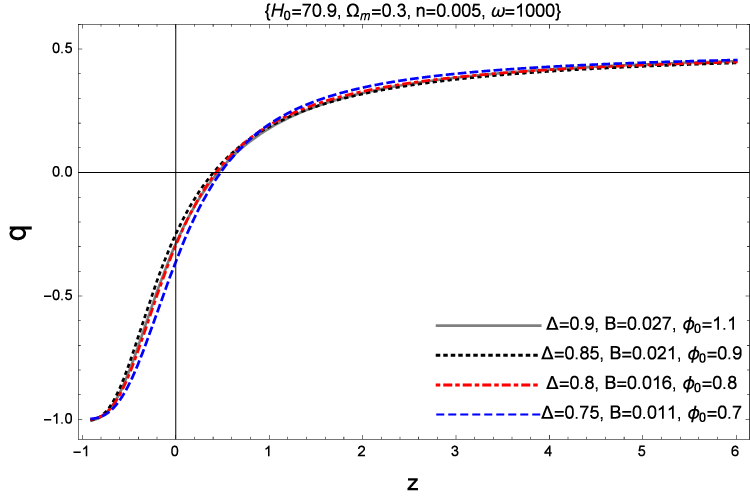}
\vspace{1mm}
\includegraphics[width=8.7cm]{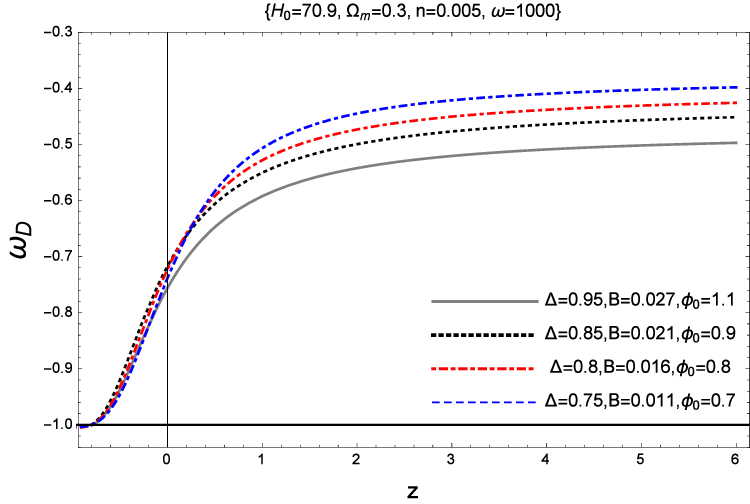}
\vspace{1mm}
\includegraphics[width=8.7cm]{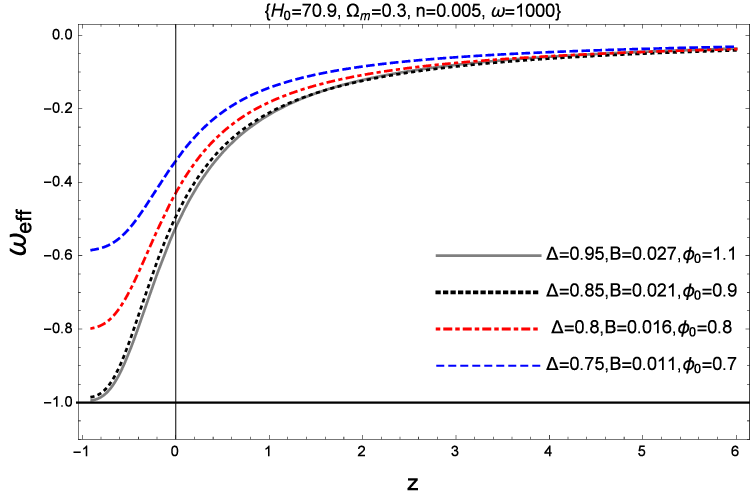}
 \caption{The evolution of the $\Omega_D(z)$, $\omega_D(z)$, $\omega_{eff}(z)$  
 and $q(z)$ parameters for non-interacting BHDE in BD gravity.} \label{fig1}
\end{center}
\end{figure}

In Fig.~\ref{fig1}, we  plot the evolution of cosmological parameters of the BHDE model  
by using a numerical solution of the Hubble differential Eq.~(\ref{Hdot1}) 
for $H_0=70.9, \Omega_m=0.3, \omega=1000$ and $n=0.05$ as the initial condition.
In the first panel of this figure, we have $\Omega_D\rightarrow 0$ at the early time 
(DM dominated phase), while at the late time the DE dominates $\Omega_D\rightarrow 1$, in agreement with cosmological observations~\cite{Daly,Komatsu,Salvatelli}.
In the second panel of Fig.~\ref{fig1}, we  show the behavior of 
the deceleration parameter $q(z)$. In contrast to standard HDE, we can see that non-interacting BHDE 
with the Hubble radius as IR cutoff can explain 
the current acceleration phase. We can see that the Universe undergoes a phase transition 
from deceleration to  acceleration around $z\approx 0.6$.

Furthermore, as it can be seen from the third panel, the EoS parameter is always 
in the quintessence regime ($\omega_D>-1$) and, at the late time 
$(z\rightarrow-1)$, it tends to the cosmological constant ($\omega_D\rightarrow -1$).

Finally, we  plot the behavior of the  total EoS parameter 
$\omega_{eff}(z)$ in the last panel of Fig.~\ref{fig1}. 
We can see that the total EoS parameter can suitably describe 
the Universe history, as the pressureless DM is dominant 
$(\omega_{eff}\rightarrow 0)$ at early times
($z\rightarrow\infty$), then it enters the quintessence regime at
the present epoch and finally, by decreasing  values of the model parameters,
tends to a value very close to $-1$ at  late times ($z\rightarrow -1$).

\section{Interacting BHDE model}
\label{Inter}
Let us now assume that dark components interact each other and one can grow at the
expense of the other. The mutual interaction between two dark components 
is initially proposed to solve the coincidence problem~\cite{Amendola,Mangano}.
Although some observations confirm the mutual interaction between dark matter and dark energy~\cite{Bertolami}, 
since the nature of DM and DE are still unknown, 
the proposed different interaction models are purely phenomenological.
So far, different models of interaction between the two dark components 
have been proposed in literatures to study the dynamics of Universe
(for details, one can look into~\cite{Quartin,Pavon,Boehmer,Caldera,Mamon,Mamon2,WeiQq1,WeiQq2,Abdollahi Zadeh, Rocco} and references therein).

In the presence of interaction between two dark components, the conservation equations are
\begin{equation}
\dot{\rho}_D+3H\left(1+\omega_D\right)\rho_D=-Q,\label{QConserveDE}
\end{equation}
\begin{equation}
\dot{\rho}_m+3H\rho_m=Q,\label{QConserveCDM}
\end{equation}
where $Q$ represents the interaction term.
Positive value of Q indicates that energy transfers from the BHDE to the DM, 
while for $Q < 0$, the reverse scenario will occur.
In the following, we consider the sign-changeable and linear models of interaction to investigate
the cosmological evolution of BHDE in the BD gravity.

\subsection{Sign-changeable Interaction}
Following~\cite{WeiQq2,Abdollahi Zadeh}, we consider  a sign-changeable interaction term as
\begin{equation}
 Q=3b^2Hq\left(\rho_m+\rho_D\right),
\end{equation}
where $b$ is the coupling constant of interaction term and $q$ is the deceleration parameter.
Taking the time derivative of Eq.~(\ref{Friedeq01}) and using Eqs.~(\ref{phidot}),
~(\ref{rhodot}) and~(\ref{QConserveCDM}), we get
 \begin{eqnarray}\label{Hdot2}
 \nonumber
&&\frac{\dot{H}}{H^2}=\Big[-\frac{9\Omega_m^0H_0^2\phi_0^2\left(1+z\right)^3}{4\omega \phi^2 H^2}+\frac{3}{2\omega}\left(2n-\frac{2\omega n^2}{3}-1\right)\\[2mm]
&&\times \left(n+\frac{3b^2}{2}\right) +
nB\left(2+\Delta	\right)\left(\frac{\phi}{H}\right)^\Delta\Big]\nonumber\\
&&\times\nonumber 
\Bigg[B\left(\Delta-2\right)\left(\frac{\phi}{H}\right)^\Delta+\left(\frac{3}{2\omega}-n^2+\frac{3n}{\omega}\right)\left(1+\frac{3b^2}{2}\right)\Bigg]^{-1}.\\
 \end{eqnarray}
Combining Eqs.~(\ref{QConserveDE}) and~(\ref{rhodot}), one can obtain the following expression for EoS parameter
\begin{eqnarray}\label{w2}
\omega_D&=&-1-\frac{(2+\Delta)n}{3}+\frac{(\Delta-2)\dot{H}}{3H^2}\nonumber\\[2mm]&+&\,b^2\left(1+\frac{\dot{H}}{H^2}\right)\left(\frac{1+2n-\frac{2\omega n^2}{3}}{\Omega_D}\right)\,.
\end{eqnarray}
In the absence of interaction term $(b^2=0)$, these equations reduce
to the corresponding relations of the previous section. 

\begin{figure}[htp]
\begin{center}
\includegraphics[width=8.7cm]{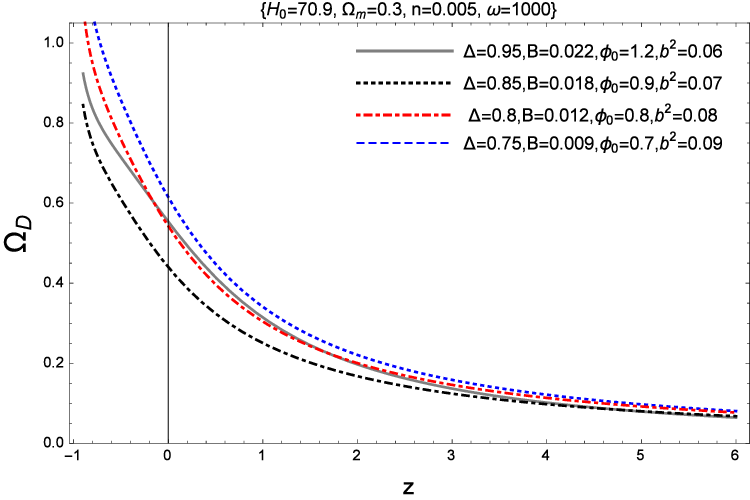}
\vspace{1mm}
\includegraphics[width=8.7cm]{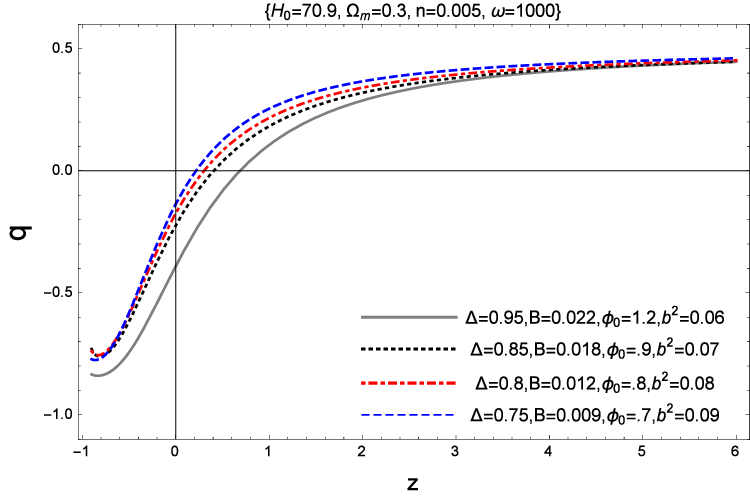}
\vspace{1mm}
\includegraphics[width=8.7cm]{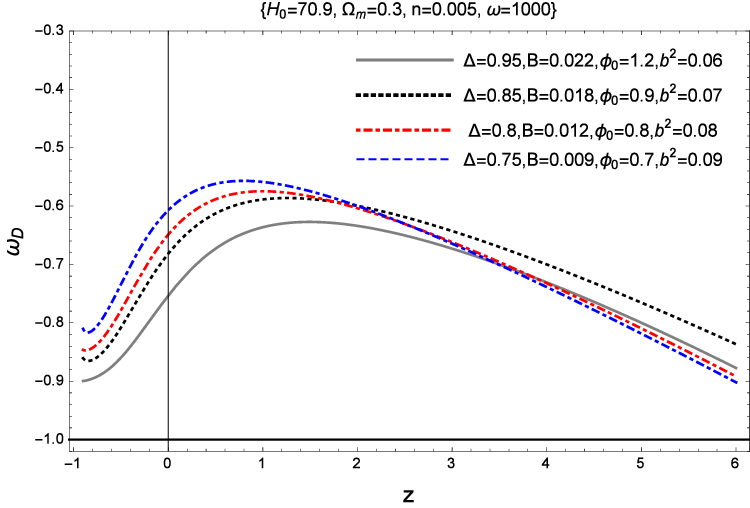}
\vspace{1mm}
\includegraphics[width=8.7cm]{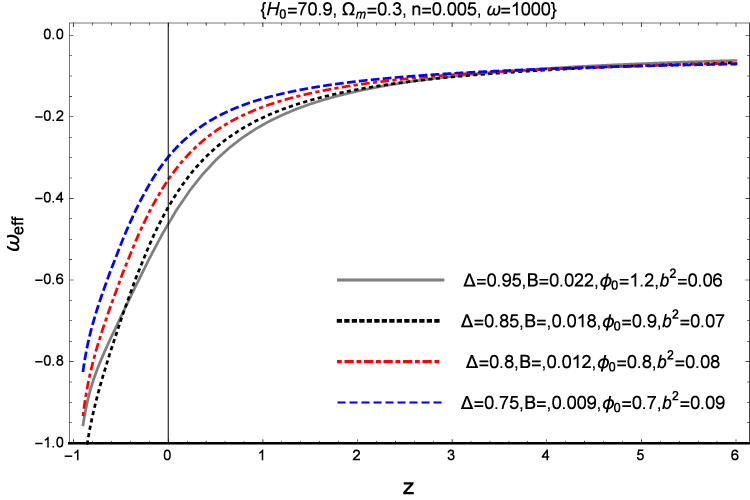}
\caption{The evolution of the $\Omega_D$, $\omega_D$, $q$ and $\omega_{eff}$ versus 
redshift parameter $z$  for sign-changeable interacting BHDE in BD gravity.
We have taken the $H_0=70.9$, $\omega_m=0.3$, $n=0.005$ and $\omega=1000$ 
as the initial condition.}\label{fig2}
\end{center}
\end{figure}

To investigate the effects of  interaction term between dark sectors,
we have plotted the behavior of $\Omega_D$, the deceleration parameter $q$,
the EoS parameter $\omega_ D$ and the total EoS parameter $\omega_{eff}$
against the redshift parameter $z$ for the sign-changeable interacting BHDE in the Fig.~\ref{fig2}.
Again, we see that at the early Universe $\Omega_D\rightarrow 0$, 
and at the late time, the DE dominates, as expected.
As it is clear from the second panel in Fiq.~\ref{fig2}, the deceleration parameter
$q$ transits from the decelerated phase $(q>0)$ at the earlier time,
to current accelerated phase $(q<0)$. Also by increasing the values of $\Delta, B$ and $\phi_0$
and decreasing the value of $b^2$, the transition occurs earlier (higher redshift).
Interestingly, we found that,  as in the non-interacting case, the EoS parameter 
of  sign-changeable interacting BHDE cannot cross the phantom divide 
$(\omega_D=-1)$ at  late time.


\subsection{Linear Interaction}
\begin{figure}[htp]
\begin{center}
\includegraphics[width=8.7cm]{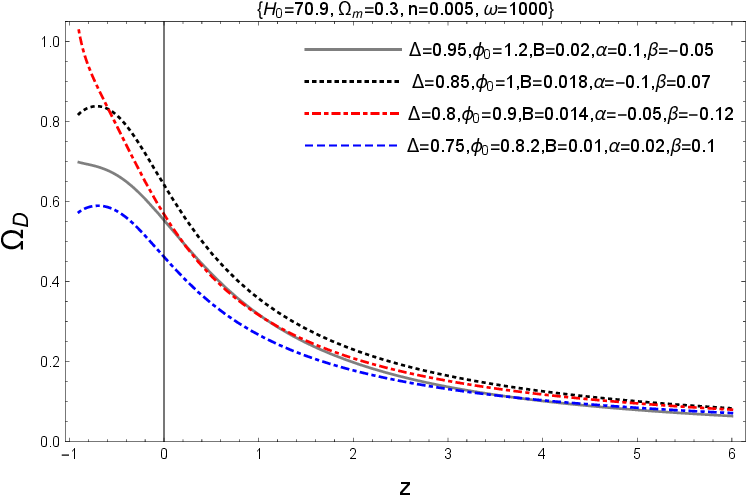}
\vspace{1mm}
\includegraphics[width=8.7cm]{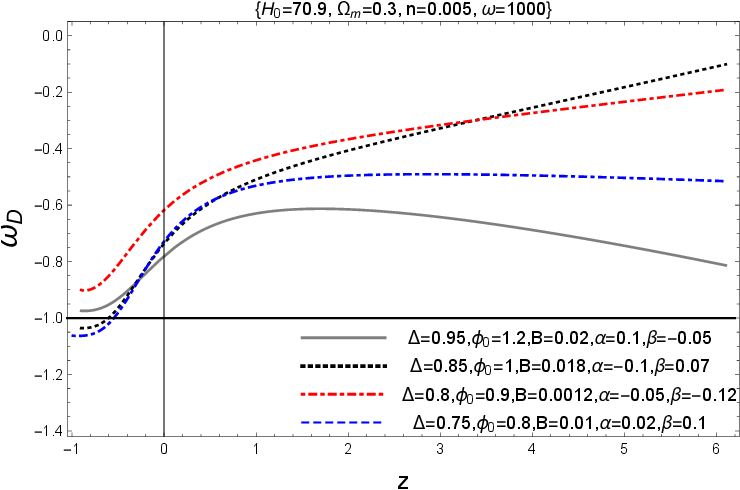}
\vspace{1mm}
\includegraphics[width=8.7cm]{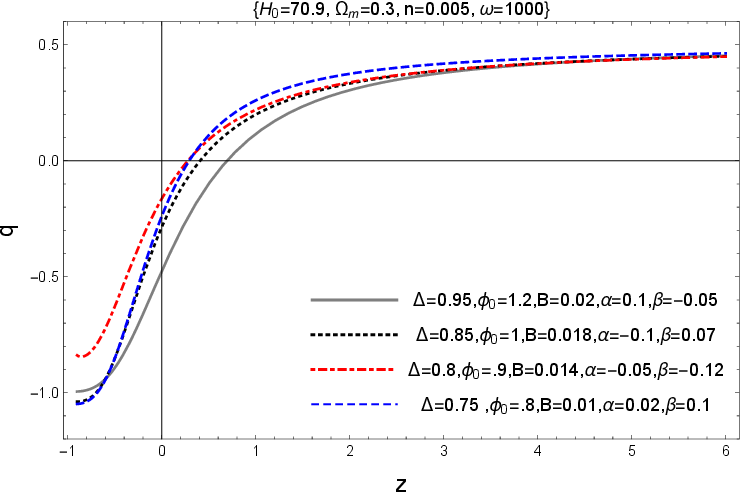}
\vspace{1mm}
\includegraphics[width=8.7cm]{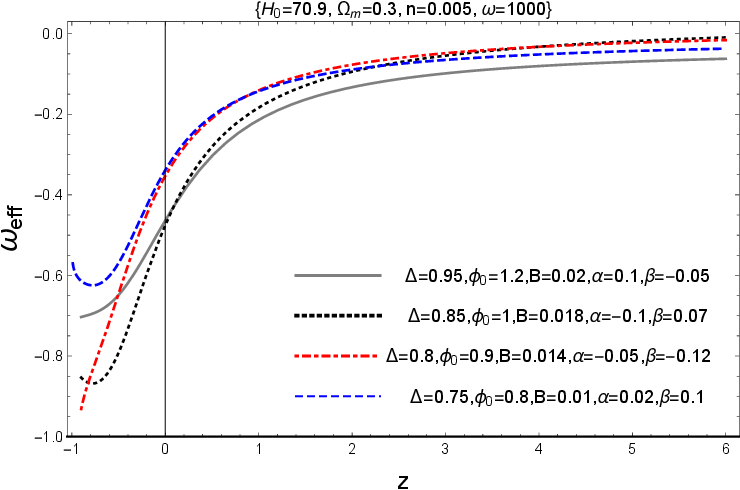}
\caption{The evolution of the $\Omega_D(z)$, $\omega_D(z)$, $q(z)$ and $\omega_{eff}(z)$ for different values 
	of $\Delta, B,\phi_0, \alpha$ and $\beta$ as mention on the caption.
for linear interacting BHDE in BD gravity.
The initial condition
$H_0=70.9$, $\omega_m=0.3$, $n=0.005$ and $\omega=1000$ are adopted.}
\label{fig3}
\end{center}
\end{figure}

In the following, we choose the linear interaction term given as~\cite{Mamon,Mamon2}
\begin{equation}\label{Q2}
Q=H(\alpha\rho_m+\beta\rho_D),
\end{equation}
where $\alpha$ and $\beta$ are the coupling constant of interaction term.
Taking the time derivative of Eq.~(\ref{Friedeq01}) and using Eqs.~(\ref{phidot}),
~(\ref{rhodot}),~(\ref{QConserveCDM}) and~(\ref{Q2}), we reach
\begin{eqnarray}\label{Hdot3}
&&\frac{\dot{H}}{H^2}=\Big[B\left(\frac{\phi}{H}\right)^\Delta(2n+\Delta n+\beta-\alpha)\nonumber\\[2mm]
&&-\,
\frac{9\Omega_m^0H_0^2\phi_0^2\left(1+z\right)^3}{4\omega \phi^2 H^2}+\frac{3}{2\omega}\left(1+2n-\frac{2\omega n^2}{3}\right)\left(\frac{\alpha}{2}-n\right)
\Big]\nonumber\\[2mm]&&\times
\Bigg[B(\Delta-2)\left(\frac{\phi}{H}\right)^\Delta+\frac{3}{2\omega}-n^2+\frac{3n}{\omega}\Bigg]^{-1}.
\end{eqnarray}
Substituting Eqs.~(\ref{rhodot}) and~(\ref{Q2}) into Eq.~(\ref{QConserveDE}), we
find out
\begin{eqnarray}\label{w3}
\omega_D&=&-1-\frac{(2+\Delta)n}{3}+\frac{(\Delta-2)\dot{H}}{3H^2}\nonumber\\[2mm]&
-&\,\frac{\alpha}{3\Omega}\left(1+2n-\frac{2\omega n^2}{3}\right)+\frac{\alpha-\beta}{3}.
\end{eqnarray}
The evolutions of $\Omega_D$, $\omega_D$, $q$ and $\omega_{eff}$ are 
plotted vs. the redshift parameter $z$ for the linear interacting BHDE in Fig.~\ref{fig3}. It appears that the linear interacting BHDE, in the BD gravity, 
can describe the cosmological evolution of the Universe in a consistent way. Indeed, from the evolution of  dimensionless BHDE density parameter $\Omega_D(z)$ in Fig.~\ref{fig3}, 
we observe that, at  early times, $\Omega_D\rightarrow 0$ and 
at  late times $\Omega_D\rightarrow 1$.
The behavior of the deceleration parameter $q(z)$ in Fig.~\ref{fig3} indicates that 
the Universe has a phase transition from deceleration to acceleration around $0.3<z<0.8$, consistently with cosmological observations~\cite{Daly,Komatsu,Salvatelli}.
The evolution of the EoS parameter $\omega_D(z)$ is shown in the third panel of Fig.~\ref{fig3}.
As we can observe, the EoS parameter successfully describes the cosmological evolution of the Universe and shows that the Universe is in the quintessence dominated phase 
($-1/3<\omega_D<-1$) at the current epoch and will enter the phantom regime
($\omega_D<-1$) in the far future.
\section{Stability Analysis}
\label{Stability}
We analyze now the  stability of the obtained models against perturbations.
So far we have investigated the dynamical behavior of  BHDE model.
Now we would like to see if there are small perturbations in the background, 
whether the disturbance grows or  collapses.
In the perturbation theory, the sign of the squared sound speed ($v_s^2 $) determines classical stability or instability.
For $v_s^2>0$, the model is stable because the perturbation  propagates on the background,
while, for $v_s^2<0$, the model is unstable since every small perturbation 
grows within the background.

The squared sound speed $v_s^2$ is given by
\begin{equation}\label{v_s}
v_s^2=\frac{dp}{d\rho_D}=\frac{\dot{p}}{\dot{\rho}_D}.
\end{equation}
By differentiating  $p_{D}$ with respect to the cosmic time,
inserting the result in Eq.~(\ref{v_s}), and using Eq.~(\ref{rhodot}), we have
\begin{equation}\label{v1}
v_s^2=\omega_D+\frac{\omega_D^\prime}{2\delta n+2(2-\delta)\frac{\dot{H}}{H^2}},
\end{equation}
for the squared sound speed.

\subsection{Non-interacting case}
\begin{figure}[htp]
\begin{center}
\includegraphics[width=8.8cm]{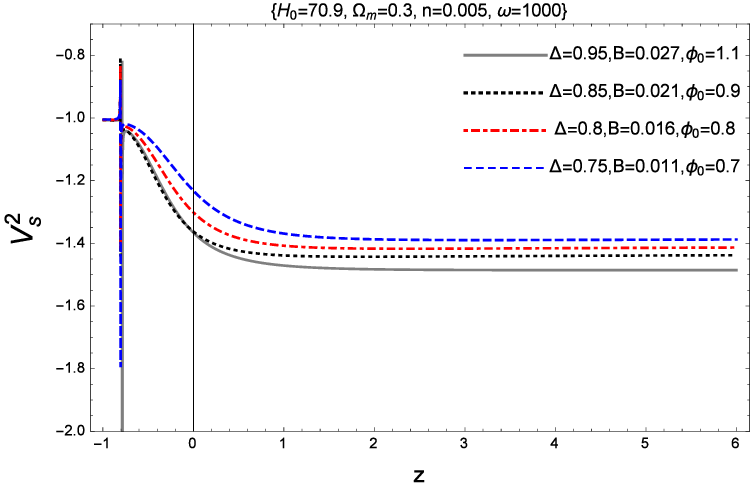}
\caption{The evolution of the $v_s^2(z)$ for non-interacting BHDE in BD gravity. 
 The initial condition
$H_0=70.9$, $\omega_m=0.3$, $n=0.005$ and $\omega=1000$ are adopted.}\label{figV1}
\end{center}
\end{figure}

Taking the time derivative of Eq.~(\ref{EoS1}) and using Eqs.~(\ref{rhodot}),~(\ref{Hdot1}) and
~(\ref{v1}), one can get $v_s^2$ for the non-interacting BHDE with the Hubble cut-off.
Since the analytic expression of $v_s^2$ is rather cumbersome to exhibit, we shall limit to plot it in Fig.~\ref{figV1}.
As  it is clear, from this figure, the non-interacting BHDE model in the BD gravity 
is unstable against  small perturbations during the cosmic evolution, which is similar to 
the result obtained for BHDE in the standard model~\cite{Srivastava}.
\subsection{Interacting case}
We now calculate the squared   sound speed for the BHDE in  presence of 
sign-changeable and linear interaction terms, respectively.
For the sign-changeable interacting case, by taking the time derivative of Eq.~(\ref{w2}) and combining the result with
Eqs.~(\ref{rho1}),~(\ref{rhodot}),~(\ref{Hdot2}) and~(\ref{v1}), we  obtain $v_s^2$.
For the linear interacting case, we use time derivative of Eq.~(\ref{w3}) and Eqs.~(\ref{rho1}),
~(\ref{rhodot}),~(\ref{Hdot3}) and~(\ref{v1}) to get $v_s^2$.
 
\begin{figure}[htp]
\begin{center}
\includegraphics[width=8.7cm]{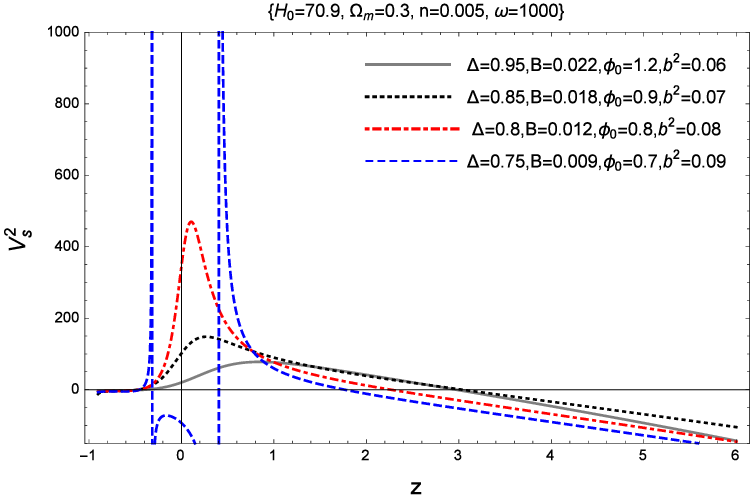}
\caption{The evolution of the $v_s^2(z)$ for sign-changeable interacting BHDE in BD gravity. 
The initial condition
$H_0=70.9$, $\omega_m=0.3$, $n=0.005$ and $\omega=1000$ are adopted.}\label{figV2}
\end{center}
\end{figure}

\begin{figure}[htp]
\begin{center}
\includegraphics[width=8.7cm]{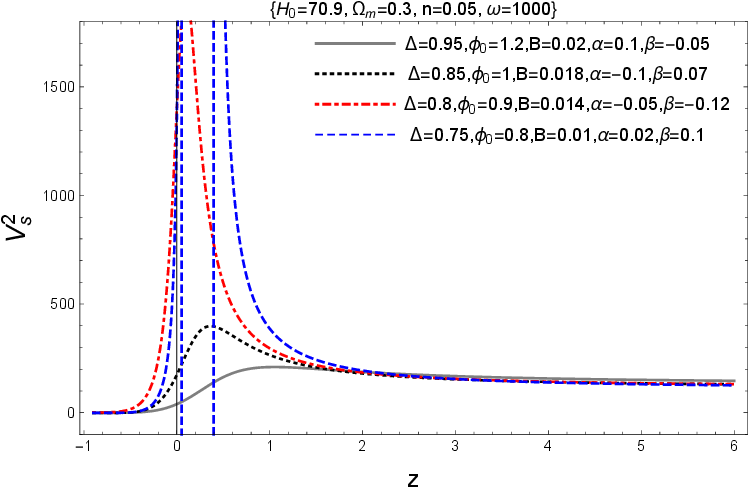}
\caption{The evolution of the $v_s^2(z)$ for linear interacting BHDE in BD gravity. 
The initial condition
$H_0=70.9$, $\omega_m=0.3$, $n=0.005$ and $\omega=1000$ are adopted.}\label{figV3}
\end{center}
\end{figure}

We  plot $v_s^2$ for the sign-changeable and linear interacting BHDE for BD gravity 
in Figs.~\ref{figV2} and~\ref{figV3}, respectively.
From Fig.~\ref{figV2}, one can clearly see that 
the sign-changeable interacting BHDE in the BD cosmology 
evolves from an unstable configuration 
in the earliest stages of the Universe's existence  
to a stable model in the present and far future. 
On the other hand, for the linear interacting BHDE 
the squared  sound speed is always positive, which means that, in this case, the Universe is always stable, thus providing a good  candidate to describe DE.

\section{Concluding remarks}
\label{Conc}
We considered the  Barrow entropy to develop a holographic DE model within the framework of Brans-Dicke
cosmology. To this goal, we assumed the Hubble horizon as the IR cutoff and studied the behavior of Barrow Holographic Dark Energy for both  interacting and non-interacting cases. We derived
the equation governing the evolution of Hubble parameter and solved it numerically in order 
to study the evolution of  corresponding cosmological parameters.

From the behavior of the deceleration parameter $q$, 
we inferred that, in contrast to the standard Holographic Dark Energy, our model can explain the present accelerated phase of the Universe expansion, even in  absence of interactions between the
dark sectors of cosmos (see~\cite{Note} for a recent discussion
on the inadequacy of standard HDE in explaining the 
accelerated expansion of the Universe consistently with
observations). 
Furthermore, for the non-interacting case,
the EoS parameter shows that BHDE model is always 
in the quintessence regime ($\omega_D>-1$) and, at the late time 
$(z\rightarrow-1)$, it tends to cosmological constant ($\omega_D\rightarrow -1$) (see Fig.~\ref{fig1}). A similar behavior is exhibited in  the sign-changeable interacting model, where the EoS  
parameter cannot cross the phantom divide at  late times (see Fig.~\ref{fig2}). 
On the other hand, in  presence of a linear interaction, the cosmological evolution of $\omega_D$ shows that 
the Universe is in the quintessence dominated phase 
at the current epoch and will enter the phantom regime
($\omega_D<-1$) in the far future (see Fig.~\ref{fig3}).

Finally, we performed a classical stability analysis using
the squared sound speed. We found that the non-interacting BHDE model, in the BD gravity, 
is unstable against  small perturbations during the cosmic evolution (see Fig.~\ref{figV1}), similarly to the result obtained for BHDE in the standard model~\cite{Srivastava}. On the other hand, the sign-changeable interacting model can be stable only for certain values of the parameters (see Fig.~\ref{figV2}), while 
the linear interacting 
model always predicts a stable Universe (see Fig.~\ref{figV3}).
This means that considering an interaction between the
dark sectors provides a more reasonable scenario
to trace the cosmological evolution  in BHDE
with BD gravity. This result can be achieved also in the framework of other interacting cosmological model \cite{Piedipalumbo}.

Further aspects remain to be addressed. For instance, 
it is interesting to examine how our model gets
modified in the more realistic scenario including also radiation
fluid in the content of the Universe. However, since
radiation effects are dominant in the early Universe only, 
we expect they do not spoil considerably our predictions 
for the current and future epochs. 
Additionally, we aim at investigating 
the correspondence between
our model and HDE based on other extensions of Boltzmann-Gibbs entropy, such as Tsallis~\cite{Ghaffari:2018,RoccoDag,LucGUP,Lucianoqgen,Lucianoqbis} and Kaniadakis~\cite{KBD} Holographic Dark Energy.  
Also, interesting results could be found
by connecting BHDE and dark energy model based on 
Gurzadyan-Xue framework. This latter scenario has been recently
addressed in~\cite{Gurza}, showing that a dark energy formula
testable with observations
can be obtained by reducing the Djorgovski-Gurzadyan integral equation~\cite{Inte} to a differential equation for the co-moving horizon.
Finally, since our 
model relies on a phenomenological attempt to include quantum 
gravitational effects through the Barrow entropy, it is important to 
study whether our predictions reconcile with
more fundamental candidate theories of quantum
gravity, such as String Theory or Loop Quantum Gravity.
Work along these directions is presently under investigation
and will be presented elsewhere. 

\acknowledgements
G. G. L. acknowledges the Spanish ``Ministerio de Universidades'' 
for the awarded Maria Zambrano fellowship and funding received
from the European Union - NextGenerationEU. 
G.G.L. and S.C. acknowledge the participation to the COST
Action CA18108  ``Quantum Gravity Phenomenology in the Multimessenger Approach''.  S.C. acknowledges the support of INFN sez. di Napoli (Iniziativa Specifica QGSKY).

\end{document}